\documentclass[11pt]{amsart}
\usepackage{amsfonts}
\usepackage{amssymb}
\newtheorem{thm}{Theorem}[section]
\newtheorem{defn}[thm]{Definition}
\newtheorem{lem}[thm]{Lemma}
\newtheorem{prop}[thm]{Proposition}
\newtheorem{cor}[thm]{Corollary}

\newtheorem{example}[thm]{Example}

\newcommand{\mb}{\mathbb}
\newcommand{\mc}{\mathcal}
\newcommand{\eul}{\mathfrak}
\newcommand{\A}{\eul A}
\newcommand{\Ao}{{\eul A}_{\scriptscriptstyle 0}}
\newcommand{\id}{{\it 1}}
\newcommand{\M}{\eul M}
\newcommand{\D}{\mc D}
\newcommand{\HH}{\mc H}

\newcommand{\hh}{\mc H}
\newcommand{\dd}{\mc D}

\newcommand{\B}{{\eul B}}
\newcommand{\Bo}{{\eul B}_{\scriptscriptstyle 0}}
\newcommand{\N}{{\eul N}}

\newcommand{\X}{{\mathfrak X}}

\newcommand{\mult}{\,{\scriptstyle \square}\,}
\newcommand{\vp}{\varphi}
\newcommand{\ip}[2]{\left\langle {#1}\left|{#2}\right.\right\rangle}

\def\x{\relax\ifmmode {\mbox{*}}\else*\fi}
\newcommand{\bedefin}{\begin{defn}$\!\!${\bf }$\;$\rm }
\newcommand{\findefin}{ \end{defn}}
\newcommand{\beex}{\begin{example}$\!\!${\bf }$\;$\rm }
\newcommand{\enex}{ \end{example}}
\newcommand{\berem}{\begin{rem}$\!\!${\bf }$\;$\rm }
\newcommand{\enrem}{ \end{rem}}

\newcommand{\SQX}{{\mb S}(\X)}

\newcommand{\up}{\raisebox{0.7mm}{$\upharpoonright \,$}}
\newcommand{\ad}{^{\scriptstyle \dag}}
\newcommand{\LDH}{{\mathcal L}\ad(\D,\hh)}

\newcommand{\LpD}{{\mathcal L}\ad(\D)}
\newcommand{\w}{_{\rm w}}
\def\id{{\it 1}}
\begin{document}
\title[Completely positive maps in partial *-algebras]{Completely positive invariant conjugate-bilinear maps in partial *-algebras}
\author{F. Bagarello}%
\address{Dipartimento di Metodi e Modelli Matematici,
Universit\`a di Palermo, Facolt\`a d'Ingegneria, I-90128 Palermo (Italy)}%
\email{bagarell@unipa.it}%
\author{A. Inoue}%
\address{Department of Applied Mathematics, Fukuoka University,
Fukuoka 814-0180 (Japan)} \email{a-inoue@fukuoka-u.ac.jp}
\author{C. Trapani}%
\address{Dipartimento di Matematica ed Applicazioni, Universit\`a di Palermo,
I-90123 Palermo (Italy)}%
\email{trapani@unipa.it}
\begin{abstract} The notion of completely positive invariant
conjugate-bilinear map in a partial *-algebra is introduced and a
generalized Stinespring theorem is proven. Applications to the
existence of integrable extensions of *-representations of
commutative, locally convex quasi*-algebras are also discussed.
\end{abstract}
\maketitle
\section{Introduction}
Completely positive linear maps on *-algebras play a relevant role
in many applications such as quantum theory, quantum information,
quantum probability theory [see \cite{sewell, fagnola}, for
overviews].

In quantum physics, for instance, these maps describe the passage
from the dynamics of a system to that of its subsystems and they
act on the observable algebra of the system itself which is
usually taken to be a C*-algebra and then represented by bounded
operators on some Hilbert space.

It is now long time that the C*-algebraic approach to quantum
theory has been shown to be a too rigid scheme to include in its
framework all objects of physical interest and several possible
generalizations have been proposed: quasi *-algebras, partial
*-algebras and so on. It is then natural to try and extend the
notion of complete positivity to these different situations that
become relevant when unbounded operators occur.

 From a mathematical point of view the most classical
result on this topic is the Stinespring dilation theorem, that
essentially says that a linear map $T:\A \to {\mathfrak B}$ where
$\A$ is a C*-algebra with unit and ${\mathfrak B}$ is a C*-algebra
of bounded operators in Hilbert space $\hh$, is completely
positive if and only if it has the form
$$T(a)= V^*\pi(a)V, \quad a\in \A$$
where $\pi$ is a bounded representation of $\A$ in Hilbert space
$\mc K$ and $V$ is a bounded linear map of $\hh$ into $\mc K$.

A more general set-up was considered by Schm\"udgen in
\cite[Ch.11]{schm_book} where he considered completely positive
maps from an arbitrary *-algebra $\A$ into a vector space $\X$ and
showed that a Stinespring-like representation holds for all
completely positive mappings of $\A$ into a vector space $\X$.
This result found applications in the study of integrable
extensions of *-representations of both  commutative *-algebras
and enveloping algebras.

This paper is devoted to the possibility of extending
Schm\"udgen's results to the case where $\A$ is a partial
*-algebra \cite{ait_book}. The lack of an everywhere defined
multiplication makes impossible to adapt the usual notion of
complete positivity for a linear map $T$, since in this case
products of the form $a^*b$, $a,b \in \A$ need not be defined. For
this reason, we consider instead of linear maps, {\em
conjugate-bilinear} maps defined on a subspace of $\A \times \A$.
But, in the same fashion as Antoine and two of us did in
\cite{ait_biweights, ait_book} for generalizing the GNS
costruction to partial *-algebras, also in this case, in order to
obtain what will be called a {\em Stinespring dilation} of the
given completely positive conjugate-bilinear map, we need to
suppose the existence of a subspace (the {\em core}) of the space
of universal right multipliers $R\A$ of $\A$ enjoying certain
conditions of {\em quasi-invariance}.

The paper is organized as follows. After giving some preliminaries
(Section 2), we prove, in Section 3, a generalized Stinespring
theorem for completely positive, conjugate bilinear,
quasi-invariant maps on a partial *-algebra $\A$, with values in a
vector space $\X$ and we examine the relationships of the related
representations when different cores are considered.

In Section 4 we consider completely positive invariant linear maps
on partial O*-algebras that  are the natural framework were
*-representations of abstract partial *-algebras are defined.

In Section 5, we discuss applications to the existence of
integrable extensions of *-representations of commutative, locally
convex quasi*-algebras.
\section{Preliminaries}
In this Section we will collect some basic definitions needed in
what follows.

\bigskip

A {\em partial *-algebra} is a complex vector space $\A$, endowed
with an involution $x \mapsto x^*$ (that is, a bijection such that
$x^{**} = x$) and a partial multiplication defined by a set
$\Gamma \subset \A \times \A$ (a binary relation) such that:

$\;$(i) $(x,y) \in \Gamma$ implies  $(y^*,x^*) \in \Gamma$;

$\,$(ii)    $(x,y_1), (x,y_2) \in \Gamma$ implies $(x, \lambda y_1
+ \mu  y_2) \in \Gamma, \, \forall \,\lambda,\mu \in {\mb C}; $

(iii) for any $(x,y) \in \Gamma$, there is defined a product $x
\cdot y \in \A$, which is     distributive w.r.  to the addition
and satisfies the relation $ (x\cdot y)^* = y^* \cdot x^*$.

\noindent We shall assume the partial *-algebra $\A$ contains a
unit $\id$, i.e., $\id^* = \id, \, (\id,x) \in \Gamma, \, \forall
\, x \in \A$, and $ \id\cdot x = x\cdot \id = x, \, \forall \, x
\in \A$. (If $\A$ has no unit, it may always be embedded into a
larger partial *-algebra with unit, in the standard fashion.)

Given the defining set $\Gamma$, spaces of multipliers are defined
in the obvious way: \begin{eqnarray*} (x,y) \in \Gamma
&\Longleftrightarrow&
      x \in L(y) \, \mbox{ or $x$ is a left multiplier of $y$ } \\
&\Longleftrightarrow&
      y\in R(x) \, \mbox{ or $y$ is a right multiplier of $x$} .
\end{eqnarray*}
A partial *-algebra $\A$ is said to be {\em semi-associative} if
$y \in R(x)$ implies $y\cdot z \in R(x)$ for every $z \in R\A$ and
$$(x \cdot y) \cdot z= x \cdot(y \cdot z).$$

Let $\A [\tau]$ be a partial *-algebra, which is a \underline{locally convex space} for the
locally convex topology $\tau$.  Then $\A [\tau]$ is called a {\em
locally convex partial *-algebra} if the following two conditions
are satisfied:

(i) the involution $x \mapsto x\x$ is $\tau$-continuous;

(ii) the maps $x \mapsto ax$ and  $x \mapsto xb$ are
$\tau$-continuous for all $a \in L\A$ and $b \in R\A$.

\smallskip
A {\em quasi *-algebra} is a couple $(\A, \A_{\scriptscriptstyle
0})$, where $\A$ is a vector space with involution $^*$,
$\A_{\scriptscriptstyle 0}$ is a *-algebra and a vector subspace
of $\A$ and $\A$ is an $\A_{\scriptscriptstyle 0}$-bimodule whose
module operations and involution extend those of
$\A_{\scriptscriptstyle 0}$ \cite{schm_book}. Of course, any quasi
*-algebra is a partial *-algebra.

\smallskip A quasi *-algebra $(\A, \Ao)$ is said to be a {\em
locally convex quasi *-algebra} if $\A$ is endowed with a locally
convex topology $\tau$ such that

(i) the involution $x \mapsto x\x$ is $\tau$-continuous;

(ii) the maps $x \mapsto ax$ and  $x \mapsto xb$ are
$\tau$-continuous, for all $a,b \in \Ao$.

(iii) $\Ao$ is $\tau$-dense in $\A$.

\smallskip
Let $\hh$ be a complex Hilbert space and $\D$ a dense subspace of
$\hh$.
 We denote by $ \LDH $ the set of all (closable) linear
operators $X$ such that $ {\D}(X) = {\D},\; {\D}(X\x) \supseteq
{\D}.$ The set $ \LDH $ is a  {\em partial *-algebra}
\cite{ait_book}
 with respect to the following operations: the usual sum $X_1 + X_2 $,
the scalar multiplication $\lambda X$, the involution $ X \mapsto
X\ad = X\x \up {\D}$ and the {\em (weak)} partial multiplication
$X_1 \mult X_2 = {X_1}\ad\x X_2$, defined by

\medskip
\begin{tabular}{l}
\hspace{1cm}$ (X_1,X_2) \in \Gamma \Leftrightarrow X_2\dd \subset
D(X_1^{\dag *}) \mbox{ and }
X_1^\dag \dd \subset D(X_2^*)$\\
\hspace{1cm}$(X_1 \mult X_2)\xi:= X_1^{\dag *}X_2 \xi, \quad
\forall \xi \in \dd.$
\end{tabular}

\medskip
If $(X_1,X_2) \in \Gamma$, we say that $X_2$ is a weak right
 multiplier of $X_1$ or, equivalently, that $X_1$ is a weak left
 multiplier of $X_2$
(we write $ X_2 \in R^{\rm w}(X_1) \; \mbox{or} \;  X_1 \in L^{\rm
w}(X_2)$).

\underline{A $\dag$-invariant subset (resp. subspace) of $\LDH$ is said to be an}  \underline{{\it O$^*$-family}} \underline{ (resp. \it{O$^*$-vector space}) on $\D$.}

A {\em partial O*-algebra}  on $\D$ is a *-subalgebra $\M$ of
$\LDH $,
 that is, $\M$ is a subspace of $\LDH $, containing the identity
and such that $X\ad \in \M\ $ whenever $X \in \M\ $ and $X_1 \mult
X_2 \in \M$ for any $X_1, X_2 \in \M$ such that $X_2 \in R^{\rm w}
(X_1).$

Let $$ \LpD = \{ X \in \LDH:\, X\D \subseteq D, \, X\ad\D
\subseteq D\}.$$ Then $\LpD$ is a *-algebra w.r.to $\mult$ and
$X_1\mult X_2 \xi= X_1(X_2\xi)$ for each $\xi \in \D$. A
*-subalgebra of $\LpD$ is called an O*-algebra \cite{schm_book}.

The following topologies on $\LDH$ will be used in this paper:
\begin{itemize}
\item the {\em weak topology }\underline{$\tau_{\rm w}^\D$}:  defined by the
seminorms $p_{\xi, \eta}$, $\xi, \eta\in \D$ where $p_{\xi,
\eta}(X)=|\ip{X\xi}{\eta}|$, $X \in \LDH$;

\item  the {\em strong topology} $\tau_s^\D$ : defined by the
seminorms $p_\xi$, $\xi\in \D$ where $p_\xi(X)=\|X\xi\|$, $X \in
\LDH$;

\item the {\em strong* topology} $\tau_{s^*}^\D$: defined by the
seminorms $p^*_\xi$, $\xi\in \D$ where \underline{$p^*_\xi(X)=\max\{\|X\xi\|,
\|X^\dag \xi\|\}$, $X \in \LDH$}.
\end{itemize}

\medskip
 A  {\em *-representation} of partial
*-algebra $\A$ is a *-homomorphism of
 $\A$ into $ {\mathcal L}\ad(\D,\hh)$, for some pair $(\D,\hh)$, $\D$ a dense subspace of $ \hh$, that is, a linear
map $\pi : \A \ \rightarrow {\mathcal L}\ad(\D,\hh)$
     such that :
(i) $\pi(a\x) = \pi(a)\ad$ for every $a \in \A$; (ii) If $a,b \in
\A$ and $a \in L(b)$ then $\pi(a) \in L^{\rm w}(\pi(b))$ and
$\pi(a) \mult \pi(b) = \pi(a b).$

If (ii) holds only when $a \in \A$ and $b \in R\A$, we say that
$\pi$ is a {\em quasi} *-representation.

If $\pi$ is a *-representation of the partial *-algebra $\A$, then
$\pi(\A)$ need not be a partial O*-algebra, but, in general, it is
only an \underline{O$^*$-vector space}.

If $\M$ is an O*-family on $\D$, the {\em graph topology} on
 $\D$ is the locally convex topology defined by the family $\{ \|
\cdot \|_X; X \in \M \}$ of seminorms: $\| \xi \|_X \equiv \| X
\xi\|, \xi \in \D$ and it is denoted by $t_\M$.

We denote by $\widetilde{\D}(\M)$ the completion of the locally
convex space $\D[t_\M]$ and put
$$
\widehat{\D} (\M) = \bigcap_{X \in \M} \D(\overline{X}).
$$
An O*-family $\M$ on $\D$ is said to be {\it closed} if $\D=
\widetilde{\D}(\M)$; and it is said to be {\it fully closed} if
$\D= \widehat{\D}(\M)$.

Now, put
$$
\D^*(\M)= \bigcap_{X \in \M} \D(X^*).$$ Then $\M$ is said to be
{\em selfadjoint} if $\D=\D^*(\M)$.

 Finally, $\M$ is said to be {\em integrable} if $\M$ is
fully closed and each $X \in \M$ such that $X=X\ad$ is essentially
selfadjoint.

\underline{The set }
$$\M'_{\sigma}=\{C \in \LDH:
\ip{X\xi}{C^*\eta}=\underline{\ip{C\xi}{X^\dag\eta}},\, \forall X\in \M,\,
\forall \xi,\eta \in \D\},$$ is called the {\em weak unbounded
commutant} of $\M$. Its bounded part $\M'_{\rm w}$ is the {\em
weak bounded commutant} of $\M$.

A fully closed partial O*-algebra $\M$ on $\D$ is called a {\it
partial GW*-algebra} if $\M'_{\rm w} \D \subset \D$ and
$\M=\M''_{{\rm w} \sigma}$.

 A *-representation $\pi$ of a partial *-algebra $\A$ is
called closed (respectively, fully closed, self-adjoint,
integrable) if $\pi(\A)$ is closed (respectively, fully closed,
self-adjoint, integrable).

\section{Generalized Stinespring theorem}
Let $\A$ be a partial *-algebra with identity $\id$ and $\X$ a
vector space. We denote with $\SQX$ the involutive vector space of
all sesquilinear forms on $\X \times \X$ with involution $\vp \to
\vp^+$ where $\vp^+(\xi, \eta)= \overline{\vp(\eta,\xi)}$, $\xi,
\eta \in \X$.

A map $\Phi:\D(\Phi)\times \D(\Phi)\to \SQX$ is said to be {\em
conjugate-bilinear} if
\begin{itemize}
    \item $\D(\Phi)$ is a subspace of $\A$;

  \item $\Phi(x,y)^+=\Phi(y,x), \quad \forall x, y \in \D(\Phi)$;
    \item$\Phi(\alpha x+\beta y,z)= \alpha \Phi(x,z)+\beta
    \Phi(y,z), \quad \forall x,y,z \in \D(\Phi),\, \forall \alpha,
    \beta \in {\mb C}$.
\end{itemize}
In particular, if $\D(\Phi)=\A$, then $\Phi$ is said to be {\em
conjugate-bilinear map on $\A \times \A$}.

It is clear that $\Phi$ is a sesquilinear map, i.e.
\begin{itemize}
    \item $\Phi(x, \alpha y+\beta z)= \overline{\alpha}
    \Phi(x,y)+\overline{\beta}
    \Phi(x,z), \quad \forall x,y,z \in \D(\Phi),\, \forall \alpha,
    \beta \in {\mb C}$.
    \end{itemize}
\bedefin \label{defn_31}A conjugate-bilinear map
$\Phi:\D(\Phi)\times \D(\Phi)\to \SQX$ is said to be {\em
quasi-invariant} if there exists a subspace $B_\Phi$ of $\D(\Phi)$
such that
\begin{itemize}
    \item[$(I)_1$:] $B_\Phi \subset R\A$;
    \item[$(I)_2$:]$\A B_\Phi\subset \D(\Phi)$;
    \item[$(I)_3$:]$\Phi(ax,y)=\Phi(x,a^*y), \quad \forall a \in
    \A, \, \forall x,y \in B_\Phi$;
    \item[$(I)_4$:]$B_\Phi$ satisfies the following density
    condition: $\forall x \in D(\Phi)$, $\forall \xi \in \X$,
    there exists a sequence $\{x_n\}\subset B_\Phi$ such that $$\lim_{n \to \infty}\Phi(x_n-x, x_n-x)(\xi,\xi)=0.$$
\end{itemize}
Furthermore, if
\begin{itemize}
    \item[$(I)'_3$]$\Phi(a^*x,by)=\Phi(x,(ab)y), \quad \forall a,b \in
    \A: a\in L(b) \, \forall x,y \in B_\Phi$
    \end{itemize}
    then $\Phi$ is said to be {\em invariant}.\\
    A subspace $B_\Phi$ satisfying the above requirements is
    called a {\em core} for $\Phi$. If $R\A$ is a core for $\Phi$,
    then $\Phi$ is said to be {\em totally invariant}.
 \findefin
In analogy with  \cite{ait_biweights, ekha1, schm_book}, we give
the following
 \bedefin \label{defn_32} A conjugate-bilinear map $\Phi:\D(\Phi)\times \D(\Phi)\to
\SQX$ is said to be {\em positive} if $\Phi(x,x)\geq 0$ (i.e.,
$\Phi(x,x)(\xi,\xi)\geq 0$ for every $\xi \in \X)$ for each $x \in
\D(\Phi)$; the map $\Phi$ is said to be {\em completely positive}
if, for each $n \in {\mb N}$,
$$ \sum_{k,l=1}^n\Phi(x_k,x_l)(\xi_k,\xi_l)\geq 0, \quad
\forall\{x_1, \ldots, x_n\}\subset \D(\Phi), \, \{\xi_1, \ldots,
\xi_n\}\subset \X.$$
 \findefin

 We now give some examples of completely positive, invariant conjugate-bilinear
 maps.
 \beex \label{ex 33} Let $\A$ be a partial *-algebra and $\X$ a
vector space. Let $\pi$ be a (quasi) *-representation of $\A$ on
the domain $\D(\pi)$. Let $V: \X \to \D(\pi)$ be a linear map. We
define a map $\Phi_{\{\pi, V\}}$ of $\A \times \A$ into $\SQX$ by
$$\Phi_{\{\pi, V\}}(a,b)(\xi, \eta)=\ip{\pi(a)V\xi}{\pi(b)V\eta},
\quad a,b \in \A,\, \xi,\eta \in \X.$$ Then $\Phi_{\{\pi, V\}}$ is
a completely positive conjugate-bilinear map on $\A \times \A$.

We put
$$B_{\{\pi, V\}}=\{x \in R\A; \pi(x)V\X \subset \D(\pi)\}.$$
If $\pi(B_{\{\pi, V\}})$ is $\tau_s^\D$-dense in $\pi(\A)$, then
$\Phi_{\{\pi, V\}}$ is (quasi-)invariant with core $B_{\{\pi, V\}}$.
 \enex
 \beex \label{ex_34}Let $\A$ be a partial *-algebra and $\pi$ a
 *-representation of $\A$. We define a map $\Phi_\pi$ of $\A\times
 \A$ into ${\mb S}(\D(\pi))$ by
 $$\Phi_\pi(a,b)(\xi, \eta)=\ip{\pi(a)\xi}{\pi(b)\eta}\quad a,b \in \A,\,
 \xi,\eta \in \D(\pi).$$

Then $\Phi_{\pi}$ is a completely positive conjugate-bilinear map
on $\A \times \A$. We put
$$B_\pi= \{x \in R\A; \pi(x)\D(\pi)\subset \D(\pi)\}.$$

If $\pi(B_{\pi})$ is $\tau_s^\D$-dense in $\pi(\A)$, then
$\Phi_{\pi}$ is invariant with core $B_{\pi}$.

Furthermore, if $\pi$ is selfadjoint, then $B_\pi=R\A$ and
$\Phi_{\pi}$ is totally invariant.
 \enex

 \beex \label{ex_35}Let $\A$ be a partial *-algebra and $\pi$ a
 (quasi) *-representation of $\A$. Let $\X$ be a vector space and
 $\A \otimes \X$ the algebraic tensor product of $\A$ and $\X$. A
 linear map $\lambda$ defined on a subspace $\D(\lambda)$ of $\A \otimes
 \X$ into $\hh_\pi$ is said to be a {\em strongly cyclic vector
 representation} of $\A \otimes \X$ for $\pi$ if there exists a
 subspace $B_\lambda$ of $\D_\lambda:=\{x\in \A; x\otimes\xi \in
 \D(\lambda),\, \forall \xi \in \X\}$ such that
 $\A B_\lambda\subset\D_\lambda,
 \pi(a)\lambda(x\otimes\xi)=\lambda(ax\otimes \xi)$ for each $a
 \in \A$, $x \in B_\lambda$ and $\xi \in \X$, and
 $\lambda(B_\lambda\otimes \X)$ is dense in $\D(\pi)[t_\pi]$.\\
 We define a map $\Phi_{\{\pi, \lambda\}}: \D_\lambda \times \D_\lambda
 \to \SQX$ by
$$\Phi_{\{\pi, \lambda\}}(x,y)(\xi,\eta)=
\ip{\lambda(x\otimes\xi)}{\lambda(y\otimes\eta)},\quad x,y \in
\D_\lambda,\, \xi, \eta \in \X.$$ Then $\Phi_{\{\pi, \lambda\}}$
is a completely positive conjugate-bilinear map on $\A \times \A$
such that
$$\Phi_{\{\pi, \lambda\}}(ax,by)(\xi,\eta)=
\ip{\pi(a)\lambda(x\otimes\xi)}{\pi(b)\lambda(y\otimes\eta)}$$ for
each $a,b \in \A,$ $x,y \in B_\lambda,$ $ \xi, \eta \in \X.$
Furthermore, if $\lambda(B_\lambda \otimes \xi)$ is dense in
$\lambda(\D_\lambda \otimes \xi)$, for each $\xi \in \X$, then
$\Phi_{\{\pi, \lambda\}}$ is (quasi-)invariant with core
$B_\lambda$.
 \enex
 \beex \label{ex_36} Let $\A[\tau]$ be a locally convex semi-associative partial
 *-algebra. Then $M\A= L\A \cap R\A$ is a *-algebra.
 Let
 $\Phi_0: M\A \to \SQX$ be a completely positive {\em linear} map
 on $M\A$. We assume that $\SQX$ is endowed with the topology $t_{{\rm S}}$ of
 simple convergence, defined by the seminorms $p_{\xi,
 \eta}(\vp)= \underline{|\Phi_0(\xi, \eta)|}$.
 We assume that
 \begin{itemize}
    \item $M\A$ is dense in $\A[\tau]$;
    \item the map $(x,y) \in M\A \times M\A \to \Phi_0(y^*x) \in
    \SQX$ is continuous with respect to the product topology
    defined by $\tau$ on $M\A$ and the topology $t_{{\rm S}}$ of
    $\SQX$.
 \end{itemize}
 For $a,b \in \A$ we define a map $\Phi$ of $\A \times \A$ into
 $\SQX$ by
 $$ \Phi(a,b)(\xi, \eta)= \lim_{\alpha,
 \beta}\Phi_0(y_\beta^*x_\alpha)(\xi, \eta), \quad \xi, \eta \in
 \X,$$
 where $\{x_\alpha\}$ and $\{y_\beta\}$ are nets in $M\A$ that
 converge to $a$ and $b$ respectively. Then $\Phi$ is a completely
 positive quasi-invariant conjugate bilinear map on $\A \times \A$
 with core $M\A$.\\
 In particular, if $\A$ is a locally convex quasi*-algebra over
 $\A_0$ (in this case $M\A=\A_0$), then $\Phi$ is a completely
 positive totally invariant conjugate bilinear map on $\A\times
 \A$ with core $\A_0$.
 \enex

 \beex \label{ex_37}
Let $\A_0[\| \cdot \|]$ be a unital C$^*$-algebra with C$^*$-norm $\| \cdot \|$ and $\tau$ a locally convex topology on $\A_0$ which is finer than the C$^*$-norm $\| \cdot \|$-topology such that $\A_0[\tau]$ is a locally convex $*$-algebra. Let $F_0$ be a completely positive linear map of $\A_0$ into the $*$-algebra $\B(\HH)$ of all bounded linear operators on a Hilbert space $\HH$. \\
(1) Suppose that the map : $(x,y) \in \Ao[\tau] \times \Ao[\tau] \mapsto F_0 (y^*x) \in \B(\HH)[\tau_{\rm w}^\D]$ is continuous for some dense subspace $\D$ in $\HH$. Then we put
\[
F(a,b) (\xi,\eta) = \lim_{\alpha, \beta} \ip{F_0 (y^*_\beta x_\alpha) \xi}{\eta}, \quad \xi, \eta \in \D,
\]
where $\{x_\alpha\} $ and $\{ y_\beta \}$ are nets in $\Ao$ which converge to $a$ and $b$ w.r.t. the topology $\tau$, respectively.
Then $F$ is a completely positive totally invariant conjugate-bilinear map of the locally convex quasi $*$-algebra $\widetilde{\Ao}[\tau]$ over $\Ao$ constructed from the completion of $\Ao[\tau]$ with core $\Ao$. \\
(2) Suppose that the map : $ x \in \Ao[\tau] \rightarrow F_0(x) \in \B(\HH)[\tau^\D_{s^*}]$ is continuous.
Then we put
\[
F(a) \xi = \lim_\alpha F_0(x_\alpha) \xi, \quad \xi \in \D,
\]
where $\{x_\alpha\}$ is a net in $\Ao$ which converges to $a$ w.r.t. $\tau$.
\\
(i) If the multiplication of $\Ao[\tau]$ is jointly continuous, then $F$ is a completely positive linear map of the locally convex $*$-algebra $\widetilde{\Ao}[\tau]$ into $\LDH$. \\
(ii) If the multiplication of $\Ao[\tau]$ is not jointly
continuous, then we can't even define the notion of complete
positivity of $F$. In this case, the results of  Section 4 can be
used.
 \enex

\beex The previous example suggests a possible physical
application concerning the time evolution of a quantum system. Let
$\Ao$ be the C*-algebra of local observables of some physical
system, in the sense of \cite{sewell}. Let $\alpha^t$ be the
automorphisms group that describes the time evolution of the
elements of $\Ao$. Then the completion $\A$ of $\Ao$ w.r. to the
{\it physical} topology \cite{lassner} is a \underline{locally convex} quasi *-algebra over
$\Ao$, which needs to be introduced because it contains physically
relevant observables as well as their time evolutions. Then, if we
define $$F_0(x,y)= \alpha^t(y^*x), \quad x,y \in \Ao, $$ $F_0$
enjoys all conditions required in the previous example, so that
the corresponding $F$ is a completely positive totally invariant
conjugate-bilinear map. \enex

 \bigskip We now show that Example \ref{ex_35} completely covers the
 general  situation; that is, for any completely positive (quasi)
 invariant conjugate bilinear map $\Phi: \D(\Phi) \times \D(\Phi)\to
 \SQX$ there exists a couple $\{\pi, \lambda\}$ consisting of a
 *-representation $\pi$ of $\A$ and of a strongly cyclic vector
 representation $\lambda$ of $\A \otimes \X$ for $\pi$ such that
 $\Phi=\Phi_{\{\pi,\lambda\}}$. This is a generalization of
 Stinespring's theorem for completely positive linear maps on von
 Neumann algebras \cite{stine}.  Generalizations of Stinespring's
 theorem have been studied by Powers \cite{powers} and Schm\"udgen
 \cite{schm_book} for O*-algebras and by Ekhaguere and Odiobala
 \cite{ekha1} and Ekhaguere \cite{ekha2} for partial *-algebras. This
 paper is aimed to generalize Schm\"udgen's
 results to partial *-algebras. The outcome is also a generalization of the studies of Ekhaguere and
 Odiobala.

 \medskip
 Let $\A$ be a partial *-algebra with identity $\id$, $\X$ a vector space and
 $\Phi$ a completely positive
 invariant conjugate bilinear map of $\D(\Phi) \times \D(\Phi)$
 into $\SQX$.
 By the complete positivity of $\Phi$, a semidefinite inner
 product $\ip{}{}$ on the algebraic tensor product $\D(\Phi)
 \otimes \X$ of $\D(\Phi)$ and $\X$ can be defined by
 $$\ip{\sum_{k=1}^n x_k\otimes \xi_k}{\sum_{l=1}^m y_l\otimes
 \eta_l}=\sum_{k=1}^n\sum_{l=1}^m\Phi(x_k,y_l)(\xi_k,\eta_l),$$
 for $\{x_k\}, \{y_l\} \subset \D(\Phi)$ and $\{\xi_k\},\{\eta_l\}
 \subset \X$.\\
 We define a subpace ${\mc N}$ of $\D(\Phi)
 \otimes \X$ by
 $${\mc N}= \left\{\sum_{k=1}^n x_k\otimes \xi_k \in \D(\Phi)
 \otimes \X; \ip{\sum_{k=1}^n x_k\otimes \xi_k}{\sum_{k=1}^n x_k\otimes \xi_k}=0
 \right\}$$
 and the coset
 $$\lambda_\Phi\left( \sum_{k=1}^n x_k\otimes \xi_k\right)= \sum_{k=1}^n x_k\otimes
 \xi_k + {\mc N}$$
 of $\sum_{k=1}^n x_k\otimes \xi_k$.\\
 Then the quotient space $\lambda_\Phi(\D(\Phi)\otimes \X)\equiv
 \D(\Phi) \otimes {\mc N}$ is a pre-Hilbert space and its completion is denoted by $\hh_\Phi$. By condition ($I_4$) of Definition
 \ref{defn_31} it is easily seen that $\lambda_\Phi(B_\Phi\otimes
 \xi)$ is dense in $\lambda_\Phi(\D(\Phi) \otimes \xi)$, for each $\xi \in \X$ and $\lambda_\Phi(B_\Phi\otimes
 \X)$ is dense in $\hh_\Phi$. We put
$$ \pi_0(a)\lambda_\Phi\left(\sum_{k=1}^n x_k\otimes \xi_k \right)=
\lambda_\Phi\left(\sum_{k=1}^n ax_k\otimes \xi_k \right)$$ for $a
\in \A $ and $\sum_{k=1}^n x_k\otimes \xi_k \in B_\Phi \otimes
\X$. Then $\pi_0$ is a *-representation of $\A$ in $\hh_\Phi$ with
$\D(\pi_0)=\lambda_\Phi(B_\Phi\otimes
 \X)$. Indeed, take arbitrary $a,b \in \A$ with $a \in \L(b)$. We
 have

 \begin{eqnarray*}\lefteqn{\ip{\pi_0(a^*)\lambda_\Phi\left(\sum_{k=1}^n x_k\otimes \xi_k \right)}
 {\pi_0(b)\lambda_\Phi\left(\sum_{l=1}^m y_l\otimes \eta_l \right)}}\\
 \hspace{1cm}&=& \ip{\lambda_\Phi\left(\sum_{k=1}^n a^*x_k\otimes \xi_k \right)}
 {\lambda_\Phi\left(\sum_{l=1}^m by_l\otimes \eta_l \right)}\\
&=&\sum_{k=1}^n\sum_{l=1}^m\Phi(a^*x_k,by_l)(\xi_k,\eta_l)\\
&=&\sum_{k=1}^n\sum_{l=1}^m\Phi(x_k,(ab)y_l)(\xi_k,\eta_l)\\
 &=&\ip{\left(\sum_{k=1}^n x_k\otimes \xi_k \right)}
 {\pi_0(ab)\lambda_\Phi\left(\sum_{l=1}^m by_l\otimes \eta_l \right)}
 \end{eqnarray*}
for each $\sum_{k=1}^n x_k\otimes \xi_k,\, \sum_{l=1}^m y_l\otimes
\eta_l \in B_\Phi\otimes \X $, which implies that $\pi_0$ is
well-defined and that it is a *-representation of $\A$. We denote
with $\pi$ its closure. Then it is clear that $\lambda_\Phi$ is a
strongly cyclic vector representation of $\A \otimes \X$ for $\pi$
with core $B_\Phi$ and that $\Phi=\Phi_{\{\pi,\lambda_\Phi\}}$. In
particular, suppose that $B_\Phi \ni \id$. We put
$$ V: \xi \in \X \to \id \otimes \xi \in B_\Phi\otimes \X .$$
Then $V$ is a linear map of $\X$ into $\D(\pi)$ such that
$\lambda_\Phi(B_\Phi\otimes \X )= \pi(B_\Phi)V\X$ and $\Phi$
equals the completely positive invariant conjugate bilinear map
$\Phi_{\{\pi,V\}}$ of Example \ref{ex 33}. The maps $\pi$ and $V$
above are denoted with $\pi_{B_\phi}$ and $V_\Phi$, respectively,
since they are determined, respectively, by the core $B_\Phi$ and
by $\Phi$ only.

In the case that $\Phi$ is quasi-invariant, $\pi_{B_\phi}$ is a
quasi *-representation of $\A$ and $\lambda_\Phi$ and $V_\Phi$ are
defined in similar way as above.

Thus we have proved the following

\begin{thm}\label{thm_37} Let $\A$ be a partial *-algebra with
identity $\id$, $\X$ a vector space and $\Phi$ a completely
positive (quasi-) invariant conjugate bilinear map of $\D(\Phi)
\otimes \D(\Phi)$ into $\SQX$. Then there exists a couple
$(\pi_{B_\Phi},\lambda_\Phi)$ consisting of a closed (quasi-)
*-representation $\pi_{B_\Phi}$ of $\A$ and a strongly cyclic
vector representation $\lambda_\Phi$ of $\A \otimes \X$ for
$\pi_{B_\Phi}$ with core $B_\Phi$ such that
$$ \Phi(ax,by)(\xi, \eta)= \ip{\pi_{B_\Phi}(a)\lambda_\Phi(x \otimes \xi)}{\pi_{B_\Phi}(b)\lambda_\Phi(y \otimes
\eta)}$$ for every $a,b \in \A$, $x,y \in B_\Phi$ and $\xi, \eta
\in \X$. In particular, if $B_\Phi \ni \id$, then there exist a
linear map $V_\Phi$ of $\X$ into $\D(\pi_{B_\Phi})$ such that
$\pi_{B_\Phi}(B_\Phi)V\X= \lambda_\Phi(B_\phi\otimes \X)$.
\end{thm}
\begin{cor}\label{cor_38} Let $\Phi$ be a completely positive totally (quasi-)
invariant conjugate-bilinear map of $\A \times \A$ into $\SQX$.
Then the couple $(\pi, V)$ of Theorem \ref{thm_37} is uniquely
determined up to unitary equivalence.
\end{cor}
\begin{proof} Let $(\rho, W)$ be another couple consisting of a
*-representation $\rho$ of $\A$ and a linear map $W$ of $\X$ into
$\D(\rho)$ such that\begin{itemize}
    \item[(i)]$\Phi(a,b)(\xi, \eta)=\ip{\rho(a)W\xi}{\rho(b)W\eta}$
    for every $a,b\in \A$ and $\xi, \eta \in \X$;
    \item[(ii)]$\rho(R\A)W\X$ is dense in $\D(\rho)[t_\rho]$.
\end{itemize}
We put
$$ U\pi(a)V\xi=\rho(a)W\xi, \quad a \in \A, \xi \in \X.$$
Then $U$ can be extended to a unitary operator of $\hh_\pi$ onto
$\hh_\rho$. We denote this extension with the same symbol $U$.
Since $\pi(R\A))V\X$ and $\rho(R\A))W\X$ are dense in
$\D(\pi)[t_\pi]$ and $\D(\rho)[t_\rho]$, respectively, it is
easily shown that $UV=W$, $U\D(\pi)=\D(\rho)$ and
$\pi(a)=U^{-1}\rho(a)U$, for each $a \in \A$. This completes the
proof.
\end{proof}
The couples $(\pi_{B_\Phi}, \lambda_\Phi)$ and $(\pi_{B_\Phi},
V_\Phi)$ for a completely positive (quasi-) invariant
conjugate-bilinear map $\Phi$ with core $B_\Phi$ are called the
{\em Stinespring dilations} of $\Phi$ determined by the core
$B_\Phi$.

In the case of a completely positive totally invariant
conjugate-bilinear map $\Phi$, $\pi_{R\A}$ is determined by $\Phi$
only and so we denote it by $\pi_\Phi$ and $(\pi_\Phi, V_\Phi)$ is
called the {\em Stinespring dilation} of $\Phi$.

\medskip
Let $\Phi$ be a a completely positive (quasi-) invariant
conjugate-bilinear map of $\D(\Phi)\times \D(\Phi)$ into $\SQX$
and denote with $\B_\Phi$ the set of all cores for $\Phi$. It may
happen that $\pi_{B_\Phi}=\pi_{B'_\Phi}$ for $B_\Phi\neq B'_\Phi$,
$B_\Phi,B'_\Phi\in \B_\Phi$. However the set of all cores that
yield the same representation has a maximal element. Indeed, we
have:

\begin{prop}\label{prop_39} Let $\Phi$ be a a completely positive (quasi-) invariant
conjugate-bilinear map of $\D(\Phi)\times \D(\Phi)$ into $\SQX$
with core $B_\Phi$. We put
\begin{eqnarray*}\lefteqn{B_\Phi^L=\left\{x \in \D(\Phi)\cap R\A;
\lambda_\Phi(x\otimes\xi) \in \D(\pi_{B_\Phi}),\, \forall \xi \in
\X;\, ax \in \D(\Phi), \right.} \\
& &
\hspace{4mm}\left.\lambda_\Phi(ax\otimes\xi)=\pi_{B_\Phi}(a)\lambda_\Phi(x\otimes
\xi),\, \forall a \in \A, \xi \in \X. \right\}.\end{eqnarray*}
Then $B_\Phi^L$ is the largest among all cores $B'_\Phi$ for which
$\pi_{B'_\Phi}=\pi_{B_\Phi}$.
\end{prop}
\begin{proof} It is easily shown that $B_\Phi^L$ is a core for
$\Phi$ such that
$$ \lambda_\Phi(B_\Phi\otimes \X) \subset
\lambda_\Phi(B_\Phi^L\otimes \X) \subset \D(\pi_{B_\Phi})$$ and
$$ \pi_{B_\Phi^L}\upharpoonright_{\lambda_\Phi(B_\Phi^L\otimes
\X)}=\pi_{B_\Phi}\upharpoonright_{\lambda_\Phi(B_\Phi^L\otimes
\X)},$$ which implies $\pi_{B_\Phi^L}=\pi_{B_\Phi}$. Take an
arbitrary core $B'_\Phi$ for $\Phi$ such that
$\pi_{B'_\Phi}=\pi_{B_\Phi}$. By the definition of $B_\Phi^L$ we
have $B_\Phi^L\supset B'_\Phi$. Thus, $B_\Phi^L$ is the largest
among the cores for $\Phi$ having the mentioned properties. This
completes the proof.
\end{proof}
We put
$$
\B_\Phi^L= \{ B_\Phi \in \B_\Phi ; B_\Phi= B_\Phi^L \}. $$

We obtain a unique characterization of a *-representation
$\pi_{B_\Phi}$ in terms of a core $B_\Phi$.

\begin{prop}

Let $\Phi$ be a a completely positive (quasi-) invariant
conjugate-bilinear map of $\D(\Phi)\times \D(\Phi)$ into $\SQX$
and $B_\Phi,B'_\Phi\in \B_\Phi$. Then the following statements
hold :

(1) $\pi_{B_\Phi} \subset \pi_{B'_\Phi}$ if and only if
$B_\Phi\subset  B'_\Phi$.

(2) $\pi_{B_\Phi} = \pi_{B'_\Phi}$ if and only if $B_\Phi=
B'_\Phi$. \end{prop}

We now specialize the generalized Stinespring theorem that we have
obtained to some particular cases. The first one is the case where
$\A$ is a locally convex quasi *-algebra. The second is the case
of completely positive totally invariant conjugate-bilinear maps
into partial O*-algebras.
\begin{cor} \label{cor_311} Let $\A$ be locally convex quasi*-algebra over $\A_0$.
Let $\Phi$ be the completely positive totally invariant
conjugate-bilinear map of $\A\times \A$ into $\SQX$ defined in
Example \ref{ex_36}.

Then the following statements hold:
\begin{itemize}
    \item[(1)]$\lambda_\Phi(\A\otimes \X)= \pi_\Phi(\A_0)V_\Phi\X$
    is dense in $\D(\pi_\Phi)[t_{\pi_\Phi}]$.
    \item[(2)]$\pi_\Phi(\A_0)$ is an O*-algebra on $\D(\pi_\Phi)$
    and $\pi_\Phi\upharpoonright_{\A_0}$ is a *-representation of
    the *-algebra $\A_0$ with
    $\D(\pi_\Phi\upharpoonright_{\A_0})\subset \D(\pi_\Phi)=
    \D(\widetilde{\pi_\Phi\upharpoonright_{\A_0}})$.
    \item[(3)]$\pi_\Phi(\A)'_{\rm w}=\pi_\Phi(\A_0)'_{\rm w}$.
\end{itemize}     \end{cor}

Let $T$ be a \underline{conjugate-bilinear} map of $\A\times \A$ into $\LDH$.
If $a,b \in \A$, we define a sesquilinear form on $\D \times \D$
by $\Phi_T(a,b)(\xi,\eta)=\ip{T(a,b)\xi}{\eta}$, $\xi,\eta \in
\D$. Then $T$ is said to completely positive if $\Phi_T$ is
completely positive. The notion of (quasi-) invariance for $T$ is
defined in similar way.

If $T$ is completely positive and totally invariant, then it
determines a couple $(\pi_{\Phi_T}, {V}_{\Phi_T})$ as described in
 Theorem \ref{thm_37}. For shortness, we put $\pi_{\Phi_T}\equiv
\pi_T$ and ${V}_{\Phi_T}=V_T$.
\begin{cor} \label{cor_312} Let $\A$ be a partial *-algebra with
identity $\id$. Let $\D$ be a dense subspace of Hilbert space
$\hh$ and $T$ a completely positive totally invariant \underline{conjugate-bilinear} map of $\A
\times \A$ into $\LDH$.
\begin{itemize}
    \item[(i)] $T(\id,\id)$ is a bounded operator if, and only if
    $\underline{\overline{V}_T}$ is a bounded linear operator of $\hh$
    into $\hh_{\pi_{T}}$.
    \item[(ii)]$T(\id,\id) =I$ if, and only if
    $\underline{\overline{V}_T}$ is an isometry of $\hh$
    into $\hh_{\pi_{T}}$.
\end{itemize}
Moreover, $T(a,\id)=V_T^*\pi_T(a)V_T, \quad \forall a\in \A$.
\end{cor}
\begin{proof} By Theorem \ref{thm_37}, we have
$$\ip{T(a,b)\xi}{\eta}=\ip{\pi_T(a)V_T\xi}{\pi_T(b)V_T\eta}, \quad
\forall a,b\in \A,\, \forall \xi, \eta \in \D.$$ Hence
$$\|V_T\xi\|^2=\ip{T(\id,\id)\xi}{\xi}, \quad \forall \xi \in
\D.$$ It is then easily shown that (i) and (ii) hold. Moreover
$$\ip{T(a,1)\xi}{\eta}=\ip{\pi_T(a)V_T\xi}{V_T\eta}=\ip{V^*\pi_T(a)V_T\xi}{\eta},
\quad \forall a \in \A ,\, \forall \xi, \eta \in \D.$$ Hence
$T(a,\id)=V_T^*\pi_T(a)V_T, \quad \forall a\in \A$.
\end{proof}

\section{Completely positive linear maps on partial O$^*$-algebras}
In this section we define and investigate completely positive invariant linear maps on partial O$^*$-algebras.

Let $\M$ be a partial O$^*$-algebra on $\D$ in $\HH$ with identity operator $I$.

\bedefin
Let $F$ be a linear map of $\M$ into $\LDH$. If there exists a completely positive conjugate-bilinear map $\overset{\circ}{F}$ of $\M \times \M$ into ${\mb S}(\D)$ such that $\overset{\circ}{F}(A,I)=F(A)$ for all $A \in \M$, then $F$ is said to be completely positive. If $\overset{\circ}{F}$ is (totally) invariant, then $F$ is said to be (totally) invariant.
\findefin

By Theorem \ref{thm_37} and Corollary \ref{cor_312} we have the generalized Stinespring theorem for completely positive invariant linear maps on partial O$^*$-algebras.

\begin{thm}\label{thm_42}
Suppose that $F$ is a completely positive \underline{totally} invariant linear map of $\M$ into $\LDH$ such that \underline{$F(I)\in \B(\HH)$ (resp. $F(I)=I$)}. Then there exists a couple $(\pi_F, V_F)$ consisting of a closed $*$-representation $\pi_F$ of $\M$ and \underline{a bounded linear map (resp. an isometry)} $V_F$ of $\D$ into $\D(\pi_F)$ such that $F(A) = V_F^* \pi_F(A) V_F$ for all $A \in \M$.
\end{thm}

We construct completely positive invariant linear maps on partial O$^*$-algebras.

\begin{prop}\label{prop_43}
Let $\M$ be a self-adjoint partial O$^*$-algebra on $\D$ in $\HH$
with identity operator $I$ and $F$ a linear map of $\M$ into
$\LDH$. Suppose that

(i) $M(\M) \equiv R^{\rm w} (\M)\ad \cap R^{\rm w}(\M)$ is
$\tau^\D_{s^*}$-dense in $\M$;

(ii) $F$ is $\tau^\D\w$-continuous;

(iii) the restriction $F \lceil_{M(\M)}$ of $F$ to the O$^*$-algebra $M(\M)$ is completely positive. \\
Then $F$ is a completely positive invariant linear map on $\M$ with core $M(\M)$.
\end{prop}
\begin{proof}
For any $A,B \in \M$ we put
\[
\overset{\circ}{F}(A,B) (\xi, \eta) = \lim_{\alpha, \beta} \ip{F(Y\ad_\beta X_\alpha) \xi}{\eta}, \quad {}^\forall \xi,\eta \in \D,
\]
where $\{ X_\alpha \}$ and $\{ Y_\beta \}$ are nets in $M(\M)$ which converge to $A$ and $B$ with respect to the topology $\tau^\D_{s^*}$, respectively.
Then it is shown that $\overset{\circ}{F}$ is a completely positive invariant conjugate-bilinear map on $\M \times \M$ with core $M(\M)$ such that $\overset{\circ}{F}(A,I)= F(A)$ for all $A \in \M$. Hence $F$ is a completely positive invariant linear map on $\M$ with core $M(\M)$.
\end{proof}

\begin{cor} \label{cor_44}
Let $\LpD_b$ be the $*$-algebra of all bounded operators in $\LpD$,
and $\M_0$ a $*$-subalgebra of $\LpD_b$ with identity operator $I$.
Suppose that $\M'_0 \D \subset \D$ and $\widetilde{\M_0}[\tau^\D_{s^*}]$ is fully closed.
Then every $\tau^\D\w$-continuous completely positive linear map $F_0$ of $\M_0$ into $\B(\HH)$ extends to a completely positive invariant linear map $F$ on the partial GW$^*$-algebra $\widetilde{\M_0}[\tau^\D_{s^*}]$ with core $\M_0$.
\end{cor}
\begin{proof}
By (\cite{ait_biweights} Corollary 2.5.13)
$\widetilde{\M_0}[\tau^\D_{s^*}]$ is a partial GW$^*$-algebra over
$\M''_0$ and $\M_0 \subset R^{\rm
w}(\widetilde{\M_0}[\tau^\D_{s^*}])\ad \cap R^{\rm
w}(\widetilde{\M_0}[\tau^\D_{s^*}])$. Since $F_0$ is $\tau^\D_{\rm
w}$-continuous, it extends to a $\tau^\D_{\rm w}$-continuous
linear map $F$ on $\widetilde{\M_0}[\tau^\D_{s^*}]$. Thus
$\widetilde{\M_0}[\tau^\D_{s^*}]$ and $F$ satisfy conditions
(i)-(iii) in Proposition \ref{prop_43}. Hence $F$ is a completely
positive invariant linear map on $\widetilde{\M_0}[\tau^\D_{s^*}]$
with core $\M_0$.
\end{proof}

 \beex \label{ex_45}
Let $\M_0$ be a von Neumann algebra on $\HH$. Let $T$ be a positive self-adjoint operator in $\HH$ affiliated with $\M_0$ and $\D^\infty(T) \equiv \displaystyle \bigcap^\infty_{n=1} \D(T^n)$.
Every $\tau^{\D^\infty(T)}_{\rm w}$-continuous completely positive linear map $F_0$ of $\M_0$ into $\B(\HH)$ extends to a completely positive invariant linear map on $\widetilde{\M_0}[\tau^{\D^\infty(T)}_{s^*}]$.
Indeed, let $T= \displaystyle \int^\infty_0 \lambda dE_T(\lambda)$ be a spectral resolution of $T$ and $\N_0$ a $*$-subalgebra generated by $I$ and $\{E_T(m) X E_T(n); m,n \in {\mb N}, X \in \M_0 \}$. Then $\N_0$ is a $*$-subalgebra of ${\mathcal L}\ad(\D^\infty(T))_b$ such that $\N'_0 = \M'_0$, $\N'_0 \D^\infty(T) \subset \D^\infty(T)$ and $\widetilde{\N_0}[\tau^{\D^\infty(T)}_{s^*}]= \widetilde{\M_0}[\tau^{\D^\infty(T)}_{s^*}]$ is fully closed. Hence it follows from Corollary \ref{cor_44} that $F_0$ extends to a completely positive invariant linear map on $\widetilde{\M_0}[\tau^{\D^\infty(T)}_{s^*}]$

In particular, every $\tau^{\D^\infty(T)}_{\rm w}$-continuous completely positive linear map of $\B(\HH)$ into $\B(\HH)$ extends to a completely positive invariant linear map on ${\mathcal L}\ad(\D^\infty(T), \HH)$.
 \enex

\section{Application to integrable extensions of *-representations of commutative
locally convex quasi *-algebras}

Let $(\A,\Ao)$ be a locally convex quasi *-algebras with unit
$\id$. Let $\tau$ be the topology of $\A$. Let also $\pi$ be a
closed *-representation of $\Ao$ which is continuous from
$\Ao[\tau]$ to $\pi(\Ao)[\tau^{\D(\pi)}_{s^*}]$. Then, for any $a\in\A$ we
put
$$ \overline\pi(a)\xi=\lim_\alpha\pi(x_\alpha)\xi, \hspace{5mm}
\xi\in\D(\pi),
$$
where $\{x_\alpha\}\subset\Ao$ is a net $\tau$-converging to $a$.
Then we have the following
\begin{lem}
$\overline\pi$ is a closed *-representation of $\A$ with
$\D(\overline\pi)=\D(\pi)$ such that:

(i) $\overline\pi(x)=\pi(x), \quad\forall x\in\Ao$;

(ii) $\overline\pi(\A)'\w=\overline\pi(\Ao)'\w$.

\begin{proof}
First of all we observe that $\overline\pi$ is a *-representation
of $\A$ and the closedness of $\pi$ implies the closedness of
$\overline{\pi}$.

(ii) In general we have $\overline\pi(\A)'\w\subset
\overline\pi(\Ao)'\w=\pi(\Ao)'\w$. Viceversa, for all
$C\in\pi(\Ao)'\w$ we have
$$\ip{C\overline\pi(a)\xi}{\eta}=\lim_\alpha\ip{C\pi(x_\alpha)\xi}{\eta}
=\lim_\alpha\ip{C\xi}{\pi(x_\alpha^*)\eta}=\ip{C\xi}{\overline\pi(a^*)\eta},$$
for all $a\in\A$ and $\xi, \eta\in\D(\overline\pi)$. Therefore $C
\in \overline\pi(\A)'\w$.
\end{proof}

\end{lem}

\vspace{2mm}

In this section we investigate under which conditions
$\overline\pi$ has an integrable extension, as an application of
the results of the previous section. In other words, we generalize
Scm\"udgen's result (\cite{schm_book}, Theorem 11.3.4), originally
given for *-algebras, to the case of partial *-algebras.

We denote by $M_n(\Bbb{C}[x_1,\ldots,x_m])$ the set of all $n
\times n$-matrices $(P_{kl}(x_1,\ldots,x_m))$ of polynomials in
the $m$ variables $x_1,\ldots,x_m$.\\
 An element $(P_{kl})$ of
$M_n(\Bbb{C}[x_1,\ldots,x_m])$ is said to be {\em positive
definite} if, for any
$(\lambda_1,\lambda_2,\ldots,\lambda_m)\in\Bbb{R}^m$, the matrix
$(P_{kl}(\lambda_1,\lambda_2,\ldots,\lambda_m))$ is positive
semi-definite, that is
$$\sum_{k,l=1}^n\,P_{kl}(\lambda_1,\lambda_2,\ldots,\lambda_m)\alpha_l\,\overline{\alpha}_k\geq
0,$$ for every $(\alpha_1,\alpha_2,\ldots,\alpha_m)\in\Bbb{C}^m$.
\\ We now put
$M(\Bbb{C}[x_1,\ldots,x_m])=\cup_{n\in\Bbb{N}}M_n(\Bbb{C}[x_1,\ldots,x_m])$.

\begin{defn}
Let $\Bo=\{b_j; \,j\in J\}$ be a subset of $(\Ao)_h=\{x\in\Ao:
x^*=x\}$ such that $\Bo\cup\{\id\}$ generates $\Ao$. Let $M(\Ao,
int)_+$ be the set of all matrices in $M(\Ao)_h$ of the form
$\left(P_{kl}(b_{j1},\ldots,b_{jm})\right)$, where $m\in\Bbb{N}$,
$(P_{kl})$ is a positive definite matrix of
$M(\Bbb{C}[x_1,\ldots,x_m])$ and $j_1,\ldots, j_m\in J$.

\end{defn}

By (\cite{schm_book}, Lemma 11.3.2), $M(\Ao,int)_+$ is independent
of $\Bo$ and it is an $m$-admissible cone in $\Ao$, that is:
\begin{itemize}
\item $M(\Ao, int)_+ + M(\Ao, int)_+\subset M(\Ao, int)_+$; \item
$\lambda M(\Ao,int)_+\subset M(\Ao, int)_+$ for all $\lambda\geq
0$; \item $M(\Ao, int)_+\cap (-M(\Ao, int)_+)=\{0\}$;\item ${\eul
P}(\Ao)\equiv\left\{\sum_{k=1}^n\x_kx^*x_k; \,x_k\in\Ao
\,(k=1,\ldots,n),\, n\in\Bbb{N}\right\}\subset M(\Ao, int)_+$ and
$x^* M(\Ao, int)_+ x\subset M(\Ao, int)_+$, $\forall x\in \Ao$.

\end{itemize}

\begin{defn}
A *-representation $\pi$ of $\Ao$ is said to be {completely
positive w.r.t. $M(\Ao, int)_+$} if the sesquilinear form
$\ip{\pi(x)\cdot}{\cdot}$ on $\D(\pi)\times\D(\pi)$ for $x\in\Ao$
is completely positive w.r.t. $M(\Ao, int)_+$, that is if
$$\ip{\sum_{k,l=1}^n(\pi(P_{kl}(b_{j1},\ldots,b_{jm}))\xi_k}{\xi_l}\geq
0$$ for each positive definite $(P_{kl}(b_{j1},\ldots,b_{jm}))\in
M_n(\Bbb{C}[x_1,\ldots,x_m])$ and \linebreak
$\{\xi_1,\ldots,\xi_n\}\subset \D(\pi)$, for each $n,m\in\Bbb{N}$.

\end{defn}

\begin{thm} \label{thm5.4}
Let $\A=\widetilde{\Ao}[\tau]$ be a commutative locally convex quasi
*-algebra with identity $\id$ and $\pi$ a closed *-representation
of the *-algebra $\Ao$ which is continuous from $\Ao[\tau]$ to
$\pi(\Ao)[\tau^{\D(\pi)}_{s^*}]$. Then the following statements are
equivalent:

(i) $\pi$ is completely positive with respect to the cone $M(\Ao,
int)_+$.

(ii) There exists an integrable *-representation of $\A$ in a
larger Hilbert space which is an extention of $\overline\pi$.

\begin{proof}
Theorem 11.3.4 of \cite{schm_book} ensures us of the existence of
an integrable *-representation $\pi_1$ in a larger Hilbert space
$\HH_1$ such that:
\begin{itemize}
\item[(5.1)] $\pi\subset \pi_1$;

\item[(5.2)] $\left(\pi_1(\Ao)'\w\right)'$ is a commutative von Neumann
algebra, see \cite{powers};

\item[(5.3)] $\pi_1(\Ao)'\w\D(\pi)$ is dense in
$\D(\pi_1)[t_{\pi_1}]$.
\end{itemize}

We put $$\rho(a)C\xi=C\overline{\pi}(a)\xi,$$ for $a\in\A$,
$C\in\pi_1(\Ao)'\w$ and $\xi\in\D(\pi)$. By (5.2) and (5.3),
$\HH_\rho$, the norm closure of $\pi_1(\Ao)'\w\D(\pi)$, equals
$\HH_{\pi_1}$. \\First, we show that $\rho$ is a *-representation
of $\A$ in $\HH_\rho=\HH_{\pi_1}$. Indeed, we have:
$$\ip{\rho(a)C\xi}{C'\eta}=\ip{C\xi}{\rho(a^*)C'\eta},\quad
\forall\,a\in\A,\,\forall\, C,\,C'\in \pi_1(\Ao)'\w,\,\forall
\xi,\,\eta\in\D(\pi).$$ This follows from the equalities
\begin{eqnarray*}\ip{\rho(a)C\xi}{C'\eta}&=&\ip{C'^*C\overline{\pi}(a)\xi}{\eta}\\&=&\lim_{\alpha}
\ip{C'^*C{\pi_1}(x_\alpha)\xi}{\eta}=\lim_{\alpha}
\ip{C'^*C\xi}{\pi_1(x_\alpha^*)\eta}\\
&=&\ip{C\xi}{C'\overline{\pi}(a^*)\eta}=\ip{C\xi}{\rho(a^*)C'\eta}
\end{eqnarray*}
Moreover $\rho(a)$ is well-defined and $\rho(a)\in{\mathcal
L}\ad(\D(\rho),\HH_{\rho})$, where
$\D(\rho)=\pi_1(\Ao)'\w\D(\pi)$.

If $a\in L(b)$ then $\rho(a)\mult\rho(b)=\rho(ab)$. Indeed, let
$C, C'\in\pi_1(\Ao)'\w$ and $\xi, \eta\in\D(\pi)$ and assume, for
the moment, that $a\in\Ao$. Then, since $\pi_1$ is integrable, we
have
\begin{eqnarray*}\ip{\rho(ab)C\xi}{C'\eta}&=&\ip{C'^*C\overline{\pi}(ab)\xi}{\eta}\\&=&\ip{C'^*C\overline{\pi}(a^*)^*
\overline{\pi}(b)\xi}{\eta}=\ip{C'^*C{\pi}(a^*)^*
\overline{\pi}(b)\xi}{\eta}\\&=& \ip{C'^*C\overline{\pi_1(a)}\,
\overline{\pi}(b)\xi}{\eta}=\ip{C'^*C
\overline{\pi}(b)\xi}{{\pi_1}(a^*)\eta}\\&=&\ip{C
\overline{\pi}(b)\xi}{C'{\pi_1}(a^*)\eta}=\ip{\rho(b)C\xi}{\rho(a^*)C'\eta}.
\end{eqnarray*}
In the case where $b\in\Ao$ the proof is slightly different. In
this case, since $\pi(b)\xi$ belongs to $\D(\pi)$ we have
\begin{eqnarray*}\ip{\rho(ab)C\xi}{C'\eta}&=&\ip{C'^*C\overline{\pi}(a^*)^*
\overline{\pi}(b)\xi}{\eta}\\&=&\ip{C'^*C\overline{\pi}(a)
{\pi}(b)\xi}{\eta}=\lim_{\alpha} \ip{C'^*C\overline{\pi}(x_\alpha)
{\pi}(b)\xi}{\eta}\\&=&\lim_{\alpha}\ip{C'^*C
\overline{\pi}(b)\xi}{{\pi}(x_\alpha^*)\eta}=\ip{C
\overline{\pi}(b)\xi}{C'\overline{\pi}(a^*)\eta}\\&=&\ip{\rho(b)C\xi}{\rho(a^*)C'\eta}.
\end{eqnarray*} Let us now prove that $\rho$ is integrable. Indeed, we can
first prove that $\pi_1(\Ao)'\w= \rho(\A)'\w$. Let, in fact, $C\in
\pi_1(\Ao)'\w$. Then, for all $a\in\A$, $C_1,\,C_2\in
\pi_1(\Ao)'\w$ and $\forall \,\xi,\,\eta\in\D(\pi)$, we have
\begin{eqnarray*}\ip{C\rho(a)C_1\xi}{C_2\eta}&=&\ip{C\,C_1\overline{\pi}(a)\xi}{C_2\eta}\\&=&\lim_{\alpha}
\ip{C\,C_1{\pi}(x_\alpha)\xi}{C_2\eta} =\lim_{\alpha}
\ip{C\,C_1\xi}{C_2{\pi}(x_\alpha^*)\eta}\\&=&\ip{C\,C_1\,\xi}{C_2\overline{\pi}(a^*)\eta}=
\ip{C\,C_1\xi}{\rho(a^*)C_2\eta}.\end{eqnarray*} Therefore
$\pi_1(\Ao)'\w \subset \rho(\A)'\w$. Conversely, take an arbitrary
$K\in \rho(\A)'\w$, $C_1,\,C_2\in \pi_1(\Ao)'\w$,
$\xi_1,\,\xi_2\in\D(\pi)$ and a generic element $x\in\Ao$ we have:
\begin{eqnarray*}\ip{K\pi_1(x)C_1\xi_1}{C_2\xi_2}&=&\ip{K\,C_1\pi_1(x)\xi_1}{C_2\xi_2}\\&=&
\ip{K\,C_1\overline{\pi}(x)\xi_1}{C_2\xi_2}=\ip{K\,\rho(x)
C_1\xi_1}{C_2\xi_2}\\&=&\ip{K\,
C_1\xi_1}{\rho(x^*)C_2\xi_2}=\ip{K\,
C_1\xi_1}{\pi_1(x^*)C_2\xi_2}.\end{eqnarray*} Since
$(\pi_1(\Ao)'\w)'\D(\pi)\subset \pi_1(\Ao)'\w\D(\pi)$ is dense in
$\D(\pi_1)[t_{\pi_1}]$, it follows that $K \in \pi_1(\Ao)'\w$. We
finally show that the closure $\tilde{\rho}$ of $\rho$ is
integrable. Indeed, the equality $(\rho(\A)'\w)'=(\pi_1(\Ao)'\w)'$
implies that $(\rho(\A)'\w)'$ is commutative and since
$\rho(\A)'\w\D(\rho)\subset \D(\rho)$, by \cite[Theorem
3.1.3]{ait_book} it follows that $\tilde{\rho}$ is integrable.

Let us now prove the converse implication: $(ii) \Rightarrow (i)$.
For this we consider an integrable *-representation $\rho$ of $\A$
in a larger Hilbert space which is an extension of
$\overline{\pi}$. Since $\pi\subset\overline{\pi}$,
$\widetilde{\rho\upharpoonright\Ao}$ is an integrable
*-representation of $\Ao$ which is an extension of $\pi$, so that,
by \cite[Theorem 11.3.4]{schm_book}, $\pi$ is completely positive
w.r.t. $M(\Ao, int)_+$. This completes the proof.
\end{proof}

\end{thm}

Let $f_0$ be a positive linear functional on $\Ao$ such that the
sesquilinear form
$$(x,y)\in(\Ao\times\Ao){\longrightarrow}
f(y^*x)\in\Bbb{C}$$ is continuous. We put
$$f(a,b)=\lim_{\lambda,\mu}f_0(y_\mu^*x_\lambda), \quad
a,b\in\A.$$ Then $f$ is a positive sesquilinear form on
$\A\times\A$, which is {\em a completely positive totally
invariant conjugate-bilinear map on $\A\times \A$ into $\Bbb{C}$}.
Then let $(\pi_f, \lambda_f)$ be the GNS-construction relative to
$f$, that is, the Stinespring dilation. By Theorem \ref{thm5.4},
we get the following

\begin{cor}
The following statements are equivalent:

(i) $\pi_f\upharpoonright\Ao$ is completely positive w.r.t. the
cone $M(\Ao, int)_+$.

(ii) There exists  an integrable *-representation of $\A$ in a
large Hilbert space $\HH$ which is an extension of $\pi_f$.
\end{cor}
\bigskip
\noindent {\large\bf Acknowledgement}

The authors acknowledge financial support of Fukuoka University
and MIUR.


\begin{thebibliography}{99}
\bibitem{ait_biweights} {\sc J.-P.
Antoine, A.Inoue} and {\sc C.Trapani}, {\it Biweights on partial
*-algebras}, {\em J. Math. Anal. Appl.} {\bf 242} (2000) 164--190
\bibitem{ait_book} {\sc J.-P.
Antoine, A.Inoue} and {\sc C.Trapani} {\it Partial *-algebras and
Their Operator Realizations}, Kluwer, Dordrecht, 2002
\bibitem{ekha1}{\sc G.O.S. Ekhaguere} and {\sc P.O. Odiobala}, Completely positive conjugate-bilinear maps
on partial $*$-algebras. {\em J. Math. Phys.} {\bf 32} (1991)
2951--2958.
\bibitem{ekha2}{\sc G. O. S. Ekhaguere}, Representation of completely positive maps
between partial *-algebras, {\em Internat. J. Theoret. Phys.} {\bf
35} (1996) 1571--1580
\bibitem{fagnola}{\sc F.Fagnola} Quantum Markov semigroups and quantum
flows, {\em Proyecciones},{\bf 18}(1999)1--144
\bibitem{powers}{\sc R. T. Powers}, Self-adjoint algebras of
unbounded operators. II, {\em Trans. Amer. Math. Soc.} {\bf 187}
(1974), 261--293.
\bibitem{lassner} G. Lassner,
{\em Topological algebras and their applications in Quantum
Statistics}, Wiss. Z. KMU-Leipzig, Math.-Naturwiss. R.,  {\bf 30}
(1981), 572--595.
\bibitem{schm_book}{\sc K. Schm\"udgen}, {\it Unbounded operator algebras
and Representation theory}, Birkh{\"a}user, Basel, 1990
\bibitem{sewell} {\sc G. L. Sewell}, {\it Quantum mechanics and its emergent macrophysics}, Princeton University Press, 2002
\bibitem{stine}{\sc W.F. Stinespring},
Positive functions on $C\sp *$-algebras. {\em Proc. Amer. Math.
Soc.} {\bf 6}, (1955)211--216.
\end{thebibliography}
\end{document}